Title: Pseudoconformal field theory at "wrong level".
//
Authors: Doron Gepner and Dmitry Kerner.
//
Abstract:
Pseudoaffine theories are characterized by formal replacement of the level by the fractional number:
$k\rightarrow\frac{k}{q}$, where $q$ is an integer coprime with $k(g+k)$ ($g$ being dual Coxeter number). 
An example of "forbidden" $q$ is considered (SU(2), $q=2$). The generalized pseudoaffine theory is
obtained. Its fusions are similar to the affine ones, number of fields in the spectrum is the integer
multiple of the number in the affine case, central charge is the integer multiple of the affine one. 
Spectra of minimal models are calculated.
//
\documentstyle[12pt]{article}
\newcommand{\beq}{\begin{equation}}
\newcommand{\eeq}{\end{equation}}
\newcommand{\beqa}{\begin{eqnarray}}
\newcommand{\eeqa}{\end{eqnarray}}
\newcommand{\di}{\partial}

\newcommand{\ber}{\begin{array}{c}}
\newcommand{\eer}{\end{array}}

\title{Pseudoconformal field theory at "wrong level"}
\author{Doron Gepner and Dmitry Kerner\\Department of Particle Physics\\Weizmann Institute of Science\\Rehovot, Israel}
\date{}
\begin{document}
\maketitle
\begin{abstract}
Pseudoaffine theories are characterized by formal replacement of the level by the fractional number:
$k\rightarrow\frac{k}{q}$, where $q$ is an integer coprime with $k(g+k)$ ($g$ being dual Coxeter number). 
An example of "forbidden" $q$ is considered (SU(2), $q=2$). The generalized pseudoaffine theory is
obtained. Its fusions are similar to the affine ones, number of fields in the spectrum is the integer
multiple of the number in the affine case, central charge is the integer multiple of the affine one. 
Spectra of minimal models are calculated.
\end{abstract}
\section{Introduction}
Classification of all the CFT's with the given fusion rules is a difficult problem. Verlinde formula
provides us with a hint \cite{gepner92}, given the fusions:
\beq
N^k_{ij}=\sum_m\frac{S_{im}S_{jm}S^\dagger_{mk}}{S_{0m}}
\eeq
first find all the possible $S$ matrices, then try to realize them as the full fledged CFT's.

The simplest case is boson on a lattice \cite{gepner98}. For $\lambda,\mu$ vectors of the lattice $M$ the S-matrix is:
\beq
S_{\lambda,\mu}=|M^*/M|^{-\frac{1}{2}}\exp(-2\pi i<\lambda,\mu>)
\eeq
If $M^*/M$ is a cyclic group $Z_{M^*/M}$ then the general solution of Verlinde equation is:
\beq  \label{S}
\ber
S_{\lambda,\mu}=|M^*/M|^{-\frac{1}{2}}\exp(-2\pi iq<\lambda,\mu>)\\
(q,k|M^*/M|)=1
\eer
\eeq
($q$ is coprime with $|M^*/M|$), it is just an automorphism of a cyclic group. The dimensions of fields in the 
corresponding "pseudobosonic" theory is\footnote{
Here k is the "level" of the U(1) boson on the lattice $M$, i.e. just $|M^*/M|$
}: $h_\lambda=\frac{q\lambda^2}{2k}$. In general $M^*/M$ is a sum of cyclic groups so one can take any 
automorphism of it:
\beq \label{Sh}
S_{\lambda,\mu}=|M^*/M|^{-\frac{1}{2}}\exp(-2\pi i<h(\lambda),\mu>)
\eeq
As was found in (\cite{gepner98},\cite{gepner99}) the pseudobosonic theories really exist, i.e. for any such automorphism one 
can find another lattice $\tilde{M}$ such that (\ref{Sh}) is its $S$ matrix.

One might ask: what if the condition on $q$ in (\ref{S}) is violated? Is there any theory that has $S$ matrix
"close to the (\ref{S})"? The natural generalization of the pseudobosonic theory is boson at level 
$q^{2n-1}\times k,~n\in~Z_+$.
Among the states in the bosonic spectrum there are states of dimension:
$$\Delta^{(q)}_{q^nm}=\frac{(q^nm)^2}{4q^{2n-1}k}=q\frac{m^2}{4k}=q\Delta^{(1)}_m$$
These are states that one naively would propose for the forbidden value of $q$. Their modular properties are as in
(\ref{S}) so that the S-matrix of (\ref{S}) is a part of the bigger S-matrix of the "generalized pseudobosonic
theory". n is a positive integer, so there are many theories with the S-matrix that has (\ref{S}) as a submatrix. 
Th s phenomenon will also occur later.
\\~

Similarly one can consider the affine Lie algebras and try to classify the theories that have the same fusion rules. 
The general solution of the Verlinde formula is \cite{gepner92}:
\beq
\ber \label{Saff}
S_{\Lambda,\Lambda'}=i^{|\Delta|}\left|\frac{M^*}{(k+g)M}\right|^{-\frac{1}{2}}\sum_{w\in
W}(-1)^w\exp(-2\pi i q\frac{<w(\Lambda+\rho),(\Lambda'+\rho)>}{(k+g)})
\\
(q,g(k+g))=1
\eer
\eeq
Such theories will be called "$q$ pseudoaffine theories" or just pseudoaffine, the $q=1$ case is the usual affine theory.
The $q$ pseudoaffine theory has central charge:$$c(q)=qc(1)=\frac{qk\rm{dim}G}{k+g}\rm{mod}~4$$
and spectrum with dimensions: 
$$h^{(q)}_\lambda=qh^{(1)}_\lambda=\frac{q<\lambda,\lambda+2\rho>}{2(k+g)}\rm{mod}~Z$$

To realize these theories (\cite{gepner98}) one can use decomposition of Hilbert space into a product 
of parafermions and free bosons (\cite{ZF},\cite{GQ},\cite{G87}) (decomposition of characters into string functions:
\beq \label{dec}
\chi^\Lambda(\tau)=\sum_\lambda C^\Lambda_\lambda(\tau)\theta_\lambda(\tau)
\eeq
where $C^\Lambda_\lambda(\tau)$ are parafermionic characters). 
(There exist many different parafermionic theories, the parafermions that appear here will be called $q=1$ parafermions).
By changing the bosonic lattice to another one with the same fusions different "pseudoaffine" theories 
are constructed. So e.g. for $q=p(k+g)+1$, $p$ is integer, we have: $c(q)-c(1)=$ integer and the
corresponding theories can be realized as a product of parafermions and pseudo-bosons with
$\tilde{q}=\tilde{p}k+1$ and $\tilde{p}$ satisfies:
$\frac{p<\Lambda,\Lambda+2\rho>}{2}=\frac{\tilde{p}\lambda^2}{2}~\rm{mod}~Z$. 
However in this way not all the theories are obtained.

To realize more theories consider the following construction:
\begin{itemize}
\item Given the parafermions from (\ref{dec}) (the $q=1$ parafermions), with the spectrum $\phi_\alpha$, conformal
dimensions $h_\alpha$, central charge $c=c(1)$ and fusion rules $N^k_{ij}$ try to find for some integer $q$ the new theory 
with the central
charge $c(q)=qc(1)$, spectrum $\phi_\alpha$ with conformal dimensions $q*h_\alpha$ and the same fusion rules $N^k_{ij}$.
If such a model exist it will be called $q$ parafermions.
\end{itemize}
Note that the idea of procedure is the same as the initial question: find the theory with the given fusions. Therefore we
go once more to the modular matrix. The parafermions of (\ref{dec}) are obtained from the coset 
$\frac{G}{U(1)^{\rm{rank(G)}}}$ thus the parafermionic modular matrix is:
\beq
S^{(\Delta,\lambda),(\Delta'\lambda')}_{\rm{parafermions}}=S^{\Delta,\Delta'}_{\rm{affine}}(\bar{S}_{\rm{boson}})^{(\lambda,\lambda')}
\eeq
Therefore the $q$ parafermions exist iff there exist the corresponding pseudoaffine and pseudobosonic theories. As follows
from (\ref{S}) and (\ref{Saff}) this occurs when $(q,g(k+g))=1$ and $(q,k|M^*/M|)=1$. Finally:
\beq \label{q}
(q,gk(k+g))=1
\eeq
In what follows this formal definition of $q$ parafermions will be used in the cases when (\ref{q}) is violated, just to
compare to the existing models with similar properties.

Having found the q-parafermions for some q one can again play with bosons, i.e. to multiply the
parafermions by bosons on different lattices but with the same fusions. In such a way the full
hierarchy of pseudoaffine theories is obtained.

A natural question appears: what happens when the condition (\ref{q}) is violated? In this case (by analogy with the
generalized pseudobosons) one can try to find
a theory that is a generalization of q-parafermions in the following sense
\begin{itemize}
\item  its central charge is the same as the central charge of q-parafermions
\item its spectrum consist of two parts: the "old" fields (that appear in the q-parafermionic spectrum) and 
the "new" fields (that do not appear there)
\item When "neglecting" the new fields, fusions of the old fields are the same as of
q-parafermions: $$\hat{N}^k_{ij}=N^k_{ij}$$
\end{itemize}
Certainly there exist a lot of such theories. A general way to realize some general pseudo 
CFT at level $\frac{k}{q}$ is to take the the $q$ times product of original CFT.
In particular here one can take a multiple tensor product of the $q=1$ parafermions
with themselves or orbifolds of this product etc. We are interested, however, in the "smallest deformation" of the
q-parafermions, i.e. the number of "new" fields in the spectrum should be as small as possible.

In this paper a particular case is considered: for $\frac{SU(2)_N}{U(1)}$ parafermions 
(the Zamolodchikov-Fateev parafermions) the $q=2$ is taken. It happens
that for each $N$ there exist a model of the central charge $c(2)=2c(1)$. Its spectrum includes that of $q=2$
parafermions ("old fields") and contains additional fields ("new fields").
Fusions of "old fields" are almost the same as for $q=2$ parafermions (in the sense of previous paragraph):
$$N^k_{ij}=\hat{N}^k_{ij}$$

The next step is to couple the "generalized pseudo-parafermions" to "generalized pseudo-bosons" previously described.
 We take the boson at level
$q^{2n-1}k,~n\in~Z_+$. Then among the states in the bosonic spectrum there are states of dimension:
$$\Delta^{(q)}_{q^nm}=\frac{(q^nm)^2}{4q^{2n-1}k}=q\frac{m^2}{4k}=q\Delta^{(1)}_m$$
These corresponds to the "old fields" in the parafermionic part and are coupled to them. Other fields in the bosonic
spectrum are coupled to the "new fields" in the parafermionic part, so that the "generalized pseudoaffine theory" is
constructed. This theory has the following properties:
\begin{itemize}
\item Its central charge is related to that of the initial affine theory: $$c(q)=qc(1)\rm{mod}~4$$
\item Its spectrum includes that of the $q$ pseudoaffine theory. Number of states in the spectrum is an integer multiple
of that in the initial affine theory.
\item Denoting fusions in the affine theory by $N^\lambda_{\mu\nu}$ and in the generalized pseudoaffine by 
$\hat{N}^\lambda_{\mu\nu}$ one has: $$N^\lambda_{\mu\nu}=\hat{N}^\lambda_{\mu\nu}$$
\end{itemize}
\section{Parafermions}
\subsection{General background}
$Z_N$ parafermions are defined as a collection of fields
$(\psi_i)_{i=1}^{N-1},~\psi_i\equiv\psi_{N-i}^\dagger$ with fusions: $[\psi_i]\times[\psi_j]=[\psi_{(i+j)\rm{mod}N}]$, of conformal
dimensions satisfying: $\Delta_k=\Delta_{N-k}$ with OPE's:
\beq
\hspace{-1cm}
\ber
\psi_i(z)\psi_j(w)\sim\frac{C_{i,j}}{(z-w)^{\Delta_i+\Delta_j-\Delta_{i+j}}}
\left(\psi_{i+j}(w)+(z-w)\frac{\Delta_{i+j}+\Delta_i-\Delta_j}{2\Delta_{i+j}}\di\psi_{i+j}(w)+\dots\right),
~i+j\neq N
\\
\psi_i(z)\psi_i^\dagger(w)\sim\frac{1}{(z-w)^{2\Delta_i}}\left(1+(z-w)^2\frac{2\Delta_i}{c}T(w)+\dots\right)
\eer
\eeq
here by $i+j$ it is meant $i+j$ mod $N$, T is the energy-momentum tensor of the parafermions, $C_{i,j}$ are the
structure constants that satisfy several constraints from Jacobi identities.

As was noticed in \cite{ZF} the monodromy condition for parafermions allows conformal dimensions of the
general form: $\Delta_i=q\frac{i(N-i)}{N}+m_i,m_i=m_{N-i}\in Z$. The $q$ parafermions are obtained when\footnote{
Several examples with nonzero $m_i$ were explored in \cite{Rav}, however no new unitary models were found.} $m_i=0$

The Zamolodchikov-Fateev parafermions \cite{ZF} are the $q=1$ parafermions: $\Delta_k=\frac{k(N-k)}{N}$. In this
case the central charge is fixed: $c=\frac{2(N-1)}{N+2}$. The parafermions are described 
by $\frac{SU(2)_N}{U(1)}$ (\cite{ZF},\cite{GQ}).  
Fields in spectrum are: $\chi^l_m, l=0,\dots,N,~m\in(-N+1,N),~m=l~\rm{mod}~2$. 
The fractional part of their conformal dimensions are 
\beq \label{dimpsi}
h=\frac{l(l+2)}{4(N+2)}-\frac{m^2}{4N}
\eeq

The $Z_N$ parafermions with $\Delta_k=2\frac{k(N-k)}{N}$ were explored partially in \cite{ZF2}. As noted
in the introduction for q=2 the fusions of the spectrum are necessarily different from the ones of 
the q=1 case. Our goal is to make them as similar to the q=1 case as possible. In particular for $N$
even there is a parafermionic field: $\chi^0_N\equiv\psi_{N/2}$ with the very specific fusions:
\beq
[\psi_{N/2}]\times[\psi_{N/2}]=[\psi_{N}]\equiv[0]~~[\psi_{N/2}]\times[\psi_i]=[\psi_{N/2+i}],
~[\psi_{N/2}]\times[\psi_{N/2+i}]=[\psi_{i}]
\eeq
It has no analogs in the q=1 case.Since this field is of integer dimension $\frac{N}{2}$ one can extend the chiral 
operator algebra. Fusions of this field with other fields:
\beq
[\chi^0_N]\times[\chi^\lambda_\nu]=[\chi^\lambda_{\nu+N}]
\eeq
show that it is local\footnote
{A field is local with respect to another if their OPE contains only integer powers of $(z-w)$. A field is local if it is local with
respect to itself.
}, and all the fields of the coset are local with respect to it, so modular invariance does not forbid any representation.
Therefore the representations of such "extended" algebra are denoted by $\chi^l_m,~0\leq l\leq N,~0\leq m\leq N$ 
for general even $N$.
We will consider the q=2 parafermions for odd N and q=2 extended algebra for even N.

The coset corresponding to the q=2 parafermions is\footnote{
As was noted in the introduction many cosets have the same central charge: $[\frac{SU(N)_1\oplus SU(N)_1}{SU(N)_2}]^2,~
\frac{SU(N)_1\oplus SU(N)_1}{SU(N)_2}\times\frac{SU(N)_2}{SO(N)_4}$ etc. We need however the spectrum
that contains spectrum of q=2 parafermions and fusions similar to the $q=1$ case.
} (\cite{schwimmer}): $\frac{SU(N)_1\oplus SU(N)_1}{SO(N)^D_4}$ for
$N\geq4$ and $\frac{SU(3)_1\oplus SU(3)_1}{SO(3)^D_8}$ for $N=3$. 
In the original paper (\cite{schwimmer}) more general models $\frac{SO(N)_k\oplus SU(N)_1}{SO(N)_{2+k}}$ were considered. 
Their central charge:
\beq
c=(N-1)(1-\frac{N(N-2)}{(k+N)(k+N-2)})
\eeq
For $k=2$ it coincides with the central charge of q=2 parafermions. In this case (due to conformal embedding: 
$\widehat{SO(N)}_2\subset\widehat{SU(N)}_1$) the chiral algebra can be extended (it corresponds to the D-modular invariant of the coset)
and we extend it to obtain the spectrum similar to that of q-parafermions. 
For $N=3$, due to specialities of $SO(N)$, the coset is: $\frac{SU(2)_4\oplus SU(3)_1}{SU(2)_8}$ (conformal embedding:
$\widehat{SU(2)}_4\subset\widehat{SU(3)}_1$). The cases of low N (N=3,4,6) are dealt with separately due to relations: 
\mbox{$SO(3)\approx SU(2)$},  \mbox{$SO(4)\approx SU(2)\oplus SU(2)$},  \mbox{$SO(6)\approx SU(4)$}.
\subsection{Technicalities}
To obtain the selection rules and field identifications of the coset $G/H$ the projection matrix is used 
(see \cite{CFT} for a concise introduction). This is the $r_h\times r_g$ matrix that projects fundamental 
weights of algebra $G$ to the fundamental weights of its subalgebra $H$. As a simple example consider the embedding: 
$SU(2)\subset SU(3)$. Take first the module [1,0] of $SU(3)$. 
This module can be decomposed into irreducible modules of $SU(2)$: [0], [1], [2]. 

The projection matrix for the decompostition $[1,0]\rightarrow[2]$ is $$P=(2,2)$$. The relation between the
generators of $SU(3)$ and $SU(2)$ is (in the Cartan-Weyl basis):
$$\ber J^\pm=2(E^{\pm\alpha_1}+E^{\pm\alpha_2})\\J^0=2(H^1+H^2)\eer$$

The case of decomposition: $[1,0]\rightarrow[1]\oplus[0]$ is realized by $P=(1,1)$.
The realization in terms of generators is:$$\ber J^\pm=\sqrt{2}E^{\pm(\alpha_1+\alpha_2)}\\J^0=H^1+H^2\eer$$

Having found the projection matrix one can decompose all the $SU(3)$ modules with respect to $SU(2)$, e.g. for
$P=(2,2)$: 
$$[0,1]\rightarrow[2]~~~[1,1]\rightarrow[4]\oplus[2]$$

To the same decompositions of $G$ modules with respect to $H$ there generally correspond many
projection matrices. More coarser object is index of embedding of $H\subset G$:
\beq
x_e=\frac{|P\theta_G|^2}{|\theta_H|^2},~~(\theta_G,~\theta_H~\rm{are~the~highest~roots~of~}G~\rm{and}~H)
\eeq
It is unique for a given decomposition, however different decompositions can have the same $x_e$. If
$\hat{H}_{k_h}\subset\hat{G}_{k_G}$ and the index of embedding is $x_e$ then $$k_H=x_e\times k_G$$

The selection rules for the field $\chi^\lambda_\nu$ of $G/H$ are:
\beq
P*\lambda-\nu\in P*Q,~~~Q~\rm{is~the~root~lattice~of}~G
\eeq
Field identification takes place when the nontrivial branching $A\rightarrow\tilde{A}$ of outer automorphism of Dynkin diagram occurs:
\beq
\forall\lambda\in G:~(A\hat\omega_0,\lambda)=(\tilde{A}\hat\omega_0,P*\lambda)~~\rm{mod}Z
\eeq
In this case the fields $\chi^\lambda_\nu$ and $\chi^{A\lambda}_{\tilde{A}\nu}$ are identified.\\
As an example consider the $SU(2)\subset SU(3)$ embedding. The outer automorphisms of SU(3) 
(\mbox{$A\hat\omega_0=\hat\omega_1$}$,~~$\mbox{$A^2\hat\omega_0=\hat\omega_2$}) has no nontrivial branching since:
$$(A\hat\omega_0,\lambda)=\frac{2\lambda_1+\lambda_2}{3},~~(A^2\hat\omega_0,\lambda)=
\frac{\lambda_1+2\lambda_2}{3}$$

The SU(2) automorphism: (\mbox{$\tilde{A}\hat\omega_0=\hat\omega_1$}) branches nontrivially:\\
\mbox{$(\tilde{A}\hat\omega_0,P\lambda)=\lambda_1+\lambda_2=0~\rm{mod}~Z=(1\hat\omega_0,\lambda)$}.\\
So in this case the fields $\chi^\lambda_\nu$ and $\chi^\lambda_{\tilde{A}\nu}$ are identified.
\subsection{$\frac{SU(3)_1\oplus SU(3)_1}{SU(2)^D_8}$} \label{fus3}
The embedding $\widehat{SU(2)}_4\subset\widehat{SU(3)}_1$ has index $x_e=4$, the projection matrix is: $P=(2,2)$. Therefore the selection rule
for the field\footnote
{Here and in the sequel the weights of algebras are given in the basis of fundamental weights: 
$\lambda=(\lambda_1,\lambda_2,\dots,\lambda_r)=\lambda_1*\hat{\omega}_1+\lambda_2*\hat{\omega}_2+\dots+\lambda_r*\hat{\omega}_r$. 
The "imaginary" weight $\lambda_0*\hat{\omega}_0$ of affine Lie algebras is omitted for brevity. In the case of 
$\widehat{SU(N)_1}$ the basic representations: $(\underbrace{0,\dots,0,1}_{m},0,\dots,0)=\hat{\omega}_m,m\geq0$ are 
denoted by $(m)$} $\chi^{\lambda,\mu_\nu}$ is: $\nu_1=0$ mod$2$\\
There is a nontrivial branching of Dynkin diagram outer automorphism: $A*\hat{\omega}_0=\hat{\omega}_1$. Corresponding to
it the field identification is $\chi^{\lambda,\mu}_{\nu_0,\nu_1}\equiv\chi^{\lambda,\mu}_{\nu_1,\nu_0}$.
The spectrum of the theory is summarized in the table. The spectrum of the q=2 parafermions is given for
comparison.\\
\begin{tabular}[!htb]{cc}
Spectrum of the q=2 $Z_3$ parafermions & Spectrum of the $\frac{SU(3)_1\oplus SU(3)_1}{SU(2)_8}$\\
\begin{tabular}{|c|c|c|}\hline
$\chi^0_{-2}~~\frac{4}{3}$   &$\chi^0_0~~0$   &  $\chi^0_{2}~~\frac{4}{3}$\\\hline
$\chi^1_{-1}~~\frac{2}{15}$   &$\chi^1_{1}~~\frac{2}{15}$&$\chi^1_{3}~~\frac{4}{5}$\\\hline
\end{tabular}
&
\begin{tabular}{|c|c|c|}\hline
$\chi^{0,0}_{0}~~0$             &  $\chi^{0,0}_2~~\frac{4}{5}$   &  $\chi^{0,0}_{4}~~\frac{2}{5}$\\\hline
$\chi^{1,0}_{0}~~\frac{4}{3}$   &  $\chi^{1,0}_2~~\frac{2}{15}$  &  $\chi^{1,0}_{4}~~\frac{11}{15}$\\\hline
$\chi^{2,0}_{0}~~\frac{4}{3}$   &  $\chi^{2,0}_2~~\frac{2}{15}$  &  $\chi^{2,0}_{4}~~\frac{11}{15}$\\\hline
$\chi^{0,1}_{0}~~\frac{4}{3}$   &  $\chi^{0,1}_2~~\frac{2}{15}$  &  $\chi^{0,1}_{4}~~\frac{11}{15}$\\\hline
$\chi^{0,2}_{0}~~\frac{4}{3}$   &  $\chi^{0,2}_2~~\frac{2}{15}$  &  $\chi^{0,2}_{4}~~\frac{11}{15}$\\\hline
$\chi^{1,1}_{0}~~\frac{2}{3}$   &  $\chi^{1,1}_2~~\frac{7}{15}$  &  $\chi^{1,1}_{4}~~\frac{1}{15}$\\\hline
$\chi^{2,1}_{0}~~\frac{2}{3}$   &  $\chi^{2,1}_2~~\frac{7}{15}$  &  $\chi^{2,1}_{4}~~\frac{1}{15}$\\\hline
$\chi^{1,2}_{0}~~\frac{2}{3}$   &  $\chi^{1,2}_2~~\frac{7}{15}$  &  $\chi^{1,2}_{4}~~\frac{1}{15}$\\\hline
$\chi^{2,2}_{0}~~\frac{2}{3}$   &  $\chi^{2,2}_2~~\frac{7}{15}$  &  $\chi^{2,2}_{4}~~\frac{1}{15}$\\\hline
\end{tabular}\\
\end{tabular}
\\
As one can see the two spectra are similar, the difference is in the additional fields of the coset. 
The fusions are "almost the same" (here and in the sequel we give only part of the fusion rules to illustrate the
situation):\\
\begin{tabular}[!htb]{cc}
Fusion of the q=2 $Z_3$ parafermions & Fusion of the $\frac{SU(3)_1\oplus SU(3)_1}{SU(2)^D_8}$\\
\begin{tabular}{|c|}\hline
$[\chi^0_2]*[\chi^0_{-2}]=[\chi^0_0]$\\\hline$[\chi^0_2]*[\chi^0_{2}]=[\chi^0_{-2}]$\\\hline
$[\chi^1_3]*[\chi^1_{3}]=[\chi^0_{0}]+[\chi^1_{3}]$\\\hline$[\chi^1_1]*[\chi^1_{1}]=[\chi^0_{-2}]+[\chi^1_{-1}]$\\\hline
\end{tabular}
&
\begin{tabular}{|c|}\hline
$[\chi^{1,0}_0]*[\chi^{2,0}_0]=[\chi^{0,0}_0]$\\\hline$[\chi^{1,0}_0]*[\chi^{1,0}_0]=[\chi^{2,0}_0]$\\\hline
$[\chi^{0,0}_2]*[\chi^{0,0}_2]=[\chi^{0,0}_0]+[\chi^{0,0}_2]+[\chi^{0,0,4}]$\\\hline
$[\chi^{1,0}_2]*[\chi^{1,0}_2]=[\chi^{2,0}_0]+[\chi^{2,0}_2]+[\chi^{2,0}_4]$\\\hline
\end{tabular}
\end{tabular}
\subsection{$\frac{SU(4)_1\oplus SU(4)_1}{SU(2)_4\oplus SU(2)_4}$} \label{fus4}
The projection matrix for each of the SU(4) factors is: $P=\left(\begin{array}{ccc}1&0&1\\1&2&1\end{array}\right)$.
The selection rules for a field $\chi^{\lambda,\mu}_{\nu^{(1)}\nu^{(2)}}$ ($\nu^{(1)}$ and $\nu^{(2)}$ corresponding to the two 
SU(2) factors) are:\\
\mbox{$\lambda_1+\lambda_3+\mu_1+\mu_3-\nu^{(1)}_1=\lambda_1+\lambda_3+\mu_1+\mu_3-\nu^{(2)}_1=0$ mod $2$}. 
Field identification arises from the nontrivial branching of SU(4) automorphism: $A\hat{\omega}_0=\hat{\omega}_2$ into the SU(2)
one: $\tilde{A}\hat{\omega}_0=\hat{\omega}_1$. It gives: $\nu^{(1)}_0\geq\nu^{(1)}_1,\nu^{(2)}_0\geq\nu^{(2)}_1$. 
The two spectra are given in the table:\\
\begin{tabular}[!htb]{cc}
$\ber\rm{Spectrum~of~the~q=2}\\Z_4~\rm{extended~parafermions}\eer$&$\ber\rm{Fractional~parts~of~dimensions}\\
\rm{of~the~}\frac{SU(4)_1\oplus SU(4)_1}{SO(4)_4}\eer$\\
\begin{tabular}{|c|c|}\hline
$\chi^0_0~~0$   &  $\chi^0_{2}~~\frac{3}{2}$\\\hline
$\chi^1_{1}~~\frac{1}{8}$&$\chi^1_{3}~~\frac{9}{8}$\\\hline
$\chi^2_0~~\frac{2}{3}$   &  $\chi^2_{2}~~\frac{1}{6}$\\\hline
\end{tabular}
&
\begin{tabular}{|c|c|c|c|}\hline
$\chi^{0,0}_{0}~~0$&$\chi^{0,0}_{0,2}~~\frac{2}{3}$&$\chi^{0,0}_{2,0}~~\frac{2}{3}$&$\chi^{0,0}_{2,2}~~\frac{1}{3}$\\\hline
$\chi^{0,1}_{1,1}~~\frac{1}{8}$&$\chi^{1,0}_{1,1}~~\frac{1}{8}$&&   \\\hline
$\chi^{0,2}_{0}~~\frac{1}{2}$&$\chi^{0,2}_{0,2}~~\frac{1}{6}$&$\chi^{0,2}_{2,0}~~\frac{1}{6}$&$\chi^{0,2}_{2,2}~~\frac{5}{6}$\\\hline
$\chi^{2,0}_{0}~~\frac{1}{2}$&$\chi^{2,0}_{0,2}~~\frac{1}{6}$&$\chi^{2,0}_{2,0}~~\frac{1}{6}$&$\chi^{2,0}_{2,2}~~\frac{5}{6}$\\\hline
$\chi^{1,1}_{0}~~\frac{3}{4}$&$\chi^{1,1}_{0,2}~~\frac{5}{12}$&$\chi^{1,1}_{2,0}~~\frac{5}{12}$&$\chi^{1,1}_{2,2}~~\frac{1}{12}$\\\hline
$\chi^{1,2}_{1,1}~~\frac{5}{8}$&$\chi^{2,1}_{1,1}~~\frac{5}{8}$   &&\\\hline
$\chi^{2,2}_{0}~~0$&$\chi^{2,2}_{0,2}~~\frac{2}{3}$&$\chi^{2,2}_{2,0}~~\frac{2}{3}$&$\chi^{2,2}_{2,2}~~\frac{1}{3}$\\\hline
$\chi^{0,3}_{1,1}~~\frac{1}{8}$&$\chi^{3,0}_{1,1}~~\frac{1}{8}$&&  \\\hline
$\chi^{3,2}_{1,1}~~\frac{5}{8}$&$\chi^{2,3}_{1,1}~~\frac{5}{8}$   &&\\\hline
\end{tabular}
\end{tabular}\\
Some of the fusion rules:\\
\begin{tabular}[!htb]{cc}
$\ber\rm{Fusions~of~the~q=2}\\Z_4~\rm{extended ~parafermions}\eer$&$\ber\rm{Fusions~of}\\\rm{of~}\frac{SU(4)_1\oplus
SU(4)_1}{SO(4)_4}\eer$
\\
\begin{tabular}{|c|}\hline $[\chi^0_2]*[\chi^0_2]=[\chi^0_0]$\\\hline$[\chi^0_2]*[\chi^1_1]=[\chi^1_3]$\\\hline
$[\chi^0_2]*[\chi^2_4]=[\chi^2_2]$\\\hline$[\chi^1_1]*[\chi^1_1]=[\chi^0_2]+[\chi^2_2]$\\\hline\end{tabular}
&
\begin{tabular}{|c|}\hline
$[\chi^{0,2}_0]*[\chi^{0,2}_0]=[\chi^{0,0}_0]$\\\hline$[\chi^{0,2}_0]*[\chi^{0,1}_0]=[\chi^{0,3}_0]$\\\hline
$[\chi^{0,2}_0]*[\chi^{0,0}_2]=[\chi^{0,2}_2]$\\\hline
$[\chi^{0,1}_{1,1}]*[\chi^{0,1}_{1,1}]=[\chi^{0,2}_{0,0}]+[\chi^{0,2}_{2,0}]+[\chi^{0,2}_{0,2}]+[\chi^{0,2}_{2,2}]$\\\hline\end{tabular}\\
\end{tabular}
\subsection{$\frac{SU(6)_1\oplus SU(6)_1}{SU(4)_4}$} \label{fus6}
The projection matrix for each SU(6) factor is:
$P=\left(\begin{array}{ccccc}0&1&0&1&0\\1&0&0&0&1\\0&1&2&1&0\end{array}\right)$.
The selection rules for a field $\chi^{\lambda,\mu}_{\nu}$:
\beq
\left(\!\ber\lambda_2+\lambda_4+\mu_2+\mu_4-\nu_1\\\lambda_1+\lambda_5+\mu_1+\mu_5-\nu_2\\\lambda_2+2\lambda_3+\lambda_4+
\mu_2+2\mu_3+\mu_4-\nu_3\eer\!\right)\!\!=\!\!\rm{Span}_Z\left(\!\left(\!\ber2\\-1\\0\eer\!\right),
\left(\!\ber-1\\2\\-1\eer\!\right),\left(\!\ber0\\-1\\2\eer\!\right)\!\right)
\eeq
Field identification arises from the nontrivial branching of each of the SU(6) automorphism: 
$A\hat{\omega}_0=\hat{\omega}_3$ into the SU(4) one: $\tilde{A}\hat{\omega}_0=\hat{\omega}_1$.
It gives: $\nu_0\geq{\rm{max}}(\nu_1,\nu_2,\nu_3)$.
The two spectra are given in the table\footnote
{Part of the coset fields is omitted, e.g. fields $\chi^{0,2}_{0,0,0}$ and $\chi^{2,0}_{0,0,0}$ or $\chi^{0,3}_{0,0,2}$ and 
$\chi^{0,3}_{2,0,0}$ have the same dimension so only one representative is written
}:\\
\begin{tabular}[!htb]{cc}
$\ber\rm{Spectrum~of~the~q=2}\\Z_6~\rm{extended~parafermions}\eer$&$\ber\rm{Fractional~parts~of~dimensions}\\
\rm{of~the~}\frac{SU(6)_1\oplus SU(6)_1}{SU(4)_4}\eer$\\
\begin{tabular}{|c|c|c|}\hline
$\chi^0_0~~0$   &  $\chi^0_{2}~~\frac{5}{3}$&$\chi^0_{4}~~\frac{8}{3}$\\\hline
$\chi^1_{1}~~\frac{5}{48}$&$\chi^1_{3}~~\frac{23}{16}$&$\chi^1_{5}~~\frac{101}{48}$\\\hline
$\chi^2_0~~\frac{1}{2}$   &  $\chi^2_{2}~~\frac{1}{6}$&$\chi^2_{4}~~\frac{7}{6}$\\\hline
$\chi^3_1~~\frac{41}{48}$   &  $\chi^3_{3}~~\frac{3}{16}$&$\chi^3_{5}~~\frac{41}{48}$\\\hline
\end{tabular}
&
\begin{tabular}{|c|c|c|c|}\hline
$\chi^{0,0}_{0,0,0}~~0$   &  $\chi^{0,0}_{0,2,0}~~\frac{1}{4}$  &$\chi^{0,0}_{1,0,1}~~\frac{1}{2}$\\\hline
$\chi^{0,1}_{0,0,2}~~\frac{41}{48}$&$\chi^{0,1}_{0,1,0}~~\frac{5}{48}$&$\chi^{0,1}_{1,1,1}~~\frac{23}{48}$\\\hline
$\chi^{0,2}_{0,0,0}~~\frac{2}{3}$&$\chi^{0,2}_{0,2,0}~~\frac{11}{12}$&$\chi^{0,2}_{1,0,1}~~\frac{1}{6}$\\\hline
$\chi^{0,3}_{0,0,2}~~\frac{3}{16}$&$\chi^{0,3}_{0,1,0}~~\frac{7}{16}$&$\chi^{0,3}_{1,1,1}~~\frac{13}{16}$\\\hline
$\chi^{1,1}_{0,0,0}~~\frac{5}{6}$&$\chi^{1,1}_{0,2,0}~~\frac{1}{12}$  &$\chi^{1,1}_{1,0,1}~~\frac{1}{3}$\\\hline
$\chi^{1,2}_{0,0,2}~~\frac{25}{48}$&$\chi^{1,2}_{0,1,0}~~\frac{37}{48}$&$\chi^{1,2}_{1,1,1}~~\frac{7}{48}$\\\hline
$\chi^{2,2}_{0,0,0}~~\frac{1}{3}$&$\chi^{2,2}_{0,2,0}~~\frac{7}{12}$&$\chi^{2,2}_{1,0,1}~~\frac{5}{6}$\\\hline
$\chi^{2,3}_{0,0,2}~~\frac{41}{48}$&$\chi^{2,3}_{0,1,0}~~\frac{5}{48}$&$\chi^{2,3}_{1,1,1}~~\frac{23}{48}$\\\hline
$\chi^{3,3}_{0,0,0}~~\frac{1}{2}$&$\chi^{3,3}_{0,2,0}~~\frac{3}{4}$&$\chi^{3,3}_{1,0,1}~~0$\\\hline
\end{tabular}
\end{tabular}
\subsection{$\frac{SU(N)_1\oplus SU(N)_1}{SO(N)_4},~N\geq5$, odd}
The projection matrix for each SU(N) factor is
\beq
P=\left(\begin{array}{cccccccc}1&0&\dots&\dots&\dots&\dots&0&1\\0&1&0&\dots&\dots&0&1&0\\
\dots&\dots&\dots&\dots&\dots&\dots&\dots&\dots\\0&\dots&0&2&2&0&\dots&0\end{array}\right)
\eeq
Selection rules for $\chi^{\lambda,\mu}_\nu$ are simply: $\nu_{\frac{N-1}{2}}=0$ mod$~2$. Field identification occurs due to SO(N)
automorphism and results in the rule: $\nu_0\geq\nu_1$.
The comarks of SO(N) for odd N are: (1,2,\dots,2,1). Representations with their conformal weights are given in the table:\\
\footnotesize
\begin{tabular}[!htb]{|@{}l@{}|@{}l@{}|@{}l@{}|@{}l@{}|}\hline
$h_{(0,\dots,0)}=0$&$h_{(\underbrace{0,\dots,1}_{m<\frac{N-1}{2}},\dots,0)}\!=\!\frac{m(N-m)}{2(N+2)}$&
$h_{(\underbrace{\underbrace{0,\dots,0,1}_{m<\frac{N-1}{2}},0\dots,0,1}_{l<\frac{N-1}{2}},0,\dots,0)}\!=\!\frac{m(N-m)+l(N-l)+2{\rm{min}}[m,l]}{2(N+2)}$
\\\hline
$h_{(0,\dots,0,2)}\!=\!\frac{N^2-1}{8(N+2)}$&$h_{(0,\dots,0,4)}=\frac{(N-1)(N+3)}{4(N+2)}$&
$h_{(\underbrace{0,\dots,0,1}_{m<\frac{N-1}{2}},0,\dots,0,2)}=\frac{\frac{N^2-1}{4}+m(N+2-m)}{2(N+2)}$\\\hline
\end{tabular}\\
\normalsize

Fractional part of conformal dimension of the field $\chi^{\lambda,\mu}_\nu$ is calculated as\footnote
{Conformal dimensions of $SU(N)_1$ representations: $h_{\lambda_m}=\frac{m(N-m)}{2N}$.
}: 
$$h^{SU(N)_1}_\lambda+h^{SU(N)_1}_\mu-h^{SO(N)_4}_\nu$$
As in the case of low N the spectrum of q=2 $Z_N$ parafermions is included in the one of the coset and coset fusion rules are just
extension of parafermionic ones.
\subsection{$\frac{SU(N)_1\oplus SU(N)_1}{SO(N)_4},~N\geq8$, even}
The projection matrix for each of the SU(N) factors is
\beq
P=\left(\begin{array}{ccccccccc}1&0&\dots&\dots&\dots&\dots&\dots&0&1\\0&1&0&\dots&\dots&\dots&0&1&0\\
\dots&\dots&\dots&\dots&\dots&\dots&\dots&\dots&\dots\\0&\dots&0&1&2&1&0&\dots&0\end{array}\right)
\eeq
Selection rules for $\chi^{\lambda,\mu}_\nu$ depend on N:
\begin{itemize}
\item\textbf{N=4k}\\$\lambda_1+\lambda_{N-1}+\mu_1+\mu_{N-1}-\nu_1=\lambda_3+\lambda_{N-3}+\mu_3+\mu_{N-3}-\nu_3=\dots=
\lambda_{\frac{N}{2}-1}+\lambda_{\frac{N}{2}+1}+\mu_{\frac{N}{2}-1}+\mu_{\frac{N}{2}+1}-\nu_{\frac{N}{2}-1}$ mod$2$\\
$\nu_{\frac{N}{2}-1}=\nu_{\frac{N}{2}}$ mod$2$
\item\textbf{N=4k+2}\\$\lambda_1+\lambda_{N-1}+\mu_1+\mu_{N-1}-\nu_1=\lambda_3+\lambda_{N-3}+\mu_3+\mu_{N-3}-\nu_3=\dots=
\lambda_{\frac{N}{2}-2}+\lambda_{\frac{N}{2}+2}+\mu_{\frac{N}{2}-2}+\mu_{\frac{N}{2}+2}-\nu_{\frac{N}{2}-2}$ mod$2$\\
$2\lambda_{\frac{N}{2}}+2\mu_{\frac{N}{2}}-\nu_{\frac{N}{2}}=-\nu_{\frac{N}{2}-1}$ mod$4$
\end{itemize}
Field identifications occur due to SO(N) automorphisms: $A\hat\omega_0=\hat\omega_1,~A\hat\omega_0=\hat\omega_{\frac{N}{2}}$ and
the even-N SU(N) automorphism: $A\hat\omega_0=\hat\omega_{\frac{N}{2}},~~(\lambda_0,\lambda_1,\dots,\lambda_{N-1})\rightarrow
(\lambda_{N-1},\lambda_{N-2},\dots,\lambda_1,\lambda_0)$. It results in the restriction: 
$\nu_0\geq$max($\nu_1,\nu_{\frac{N}{2}-1},\nu_{\frac{N}{2}}$).
The comarks of SO(N) for even N are: (1,2,\dots,2,1,1). Representations with their conformal weights
are given in the table:\\
\footnotesize
\begin{tabular}[!htb]{|@{}c@{}|@{}c@{}|@{}c@{}|@{}c@{}|}\hline
$h_{(0,\dots,0)}=0$&$h_{(\underbrace{0,\dots,1}_{m\leq\frac{N}{2}-2},\dots,0)}\!=\!\frac{m(N-1)}{2(N+2)}$&
$h_{(\underbrace{\underbrace{0,\dots,0,1}_{m\leq\frac{N}{2}-2},0\dots,0,1}_{l\leq\frac{N}{2}-2},0,\dots,0)}\!=
\!\frac{(m+l)(N-1)+2{\rm{min}}(m,l)}{2(N+2)}$\\\hline
$h_{(2,0,\dots,0)}\!=\!\frac{N}{N+2}$&$h_{(0,\dots,0,1,1)}=\frac{N-2}{8}$&
$h_{(0,\dots,0,2)}\!=\!h_{(0,\dots,2,0)}\!=\!\frac{N^2}{8(N+2)}$\\\hline
\end{tabular}\\
\normalsize
As in the case of even N the spectrum of q=2 $Z_N$ extended parafermions is included in the one of the coset and 
coset fusion rules are just extension of parafermionic ones.
\section{Generalized pseudoaffine theory}
In the previous section we obtained q=2 parafermions with fusions similar to the ones needed to build 
the pseudoaffine theory.
To obtain extended algebra similar to the pseudoaffine one we should multiply the parafermionic theory with the U(1) boson 
at suitable level.
For N-odd, the level is 2N, while for N-even, the N/2 level should be taken. Consider this in details:
\begin{center}The odd-N case\end{center}
In the spirit of decomposition (\ref{dec}) first the products $C^0_\lambda\theta_\lambda$ are considered, since they 
form the extended algebra. As is seen from the spectrum table, the fields of N=3 coset that have the same 
conformal dimensions as $q$-parafermions ($C^0_\lambda$) are $\chi^{0,1}_0,\chi^{0,2}_0,\chi^{1,0}_0,\chi^{2,0}_0$. 
In fact for general N the fields that correspond to $C^0_\lambda$ belong to the set: $\chi^{0,n}_0,\chi^{n,0}_0$. 
In particular the fields that have the same conformal dimension as the "generating $q$ parafermion" ($\psi_1$) are
: $\chi^{0,2}_0,\chi^{0,N-2}_0,\chi^{2,0}_0,\chi^{N-2,0}_0$.

Let's choose $\chi^{0,2}_0$ as $\psi_1$, then by taking products with itself one has\footnote
{The fusions of $SU(N)_1$ are just: $[\lambda_m]\times[\lambda_n]=[\lambda_{(m+n)\rm{mod}N}]$
}\\ \mbox{$\psi_n\sim[\chi^{0,2}_0]^n=\chi^{0,(2n\rm{mod}N)}_0$.}
Among the possible representations of U(1) at level 2N one seeks for those that complete the dimensions of $\chi^{0,2}_0$ 
to integer. This condition: $\frac{2(N-1)}{N}+\frac{m^2}{4*2N}\in Z,~m\in[-2N+1,2N]$ has only one root: $m=4$. So the
extended algebra contains fields: $\chi^{0,2n}_0\theta^{(2N)}_{4n}$.

The next step is to combine different representations of the product into irreducible representations of extended algebra and to
discard representations that are nonlocal with respect to the extended algebra. It is sufficient to check the fusions with the generating operator:
$\chi^{0,2}_0\theta^{(2N)}_4$:
\beq
[\chi^{0,2}_0\theta^{(2N)}_4]\times[\chi^{\lambda,\mu}_\nu\theta^{(2N)}_m]=[\chi^{\lambda,(\mu+2)}_\nu\theta^{(2N)}_{(m+4)}]
\eeq
So the condition of locality is: $m=2\mu\rm{mod}N$.\\
Consider now the spectrum of the pseudoaffine algebra. Since $N$ is odd any representation of the extended algebra can be
denoted by the fields of the form: $\chi^{m,0}_\nu\theta^{(2N)}_{n*N},n=0,\pm1,2$. Here $m$ is SU(N) representation
with $h_m=\frac{m(N-m)}{N}$. $\theta^{(2N)}_{n*N}$ has conformal weights: $0,\frac{1}{8N},\frac{1}{2N}$ for $n=0,\pm1,2$
respectively. The $\nu$ is a representation of $SO(N)_4$ allowed by selection rules and field identifications. 
From this data the conformal dimension of any representation of extended algebra can be obtained.
Finally, the characters of the extended algebra are:
\beq \label{char}
\sum_m\chi^{\lambda,2m}_\nu\theta^{(2N)}_{4m+n*N}, n=0,\pm1,2
\eeq
It is instructive to calculate the number of primaries of the extended algebra: since $\lambda$ in (\ref{char})
gets $N+1$ values this number is: $(N+1)\times4\times$(number of "possible $\nu$'s"). The later is 
$\frac{(N+1)(N+3)}{8}$ for $N\geq7$ odd with the only exception for $N=5$ where it is 4. So the number
of fields is an integer multiple of that of the affine case. One can define the "index"= 
$\frac{\rm{Number~of~fields~in~the~spectrum~of~pseudoaffine~theory~}}{\rm{Number~of~fields~in~the~spectrum~of~affine~theory~}}$
For $N$ odd the index is $\frac{(N+1)(N+3)}{2}$ and 16 for $N=5$.
\begin{center}The even-N case\end{center}.

As in the case of odd N, the "generating" parafermion is: $\chi^{0,2}_0$, and the boson that completes its dimension 
to integer is $\theta^{(\frac{N}{2})}_2$. The check of representations local to this algebra gives: 
\mbox{$\chi^{\lambda\mu}_\nu\theta^{(\frac{N}{2})}_m$}
is local provided: $\mu=m$ mod$N$, that in our case \mbox{($m\in -\frac{N}{2}+1\dots\frac{N}{2}$},
\mbox{$\mu=0\dots N-1$)} means just $\mu=m$.\\
The spectrum of the extended algebra is calculated similarly to the odd-N case, the only difference is in the $SO(N)$
part.
Finally the characters
are:
\beq
\sum_m\chi^{\lambda(\mu+2m)}_\nu\theta^{(\frac{N}{2})}_{\mu+2m}
\eeq
Similarly to the $N$ odd case the number of fields in the spectrum:\\ 
\mbox{$(N+1)\left(\frac{(\frac{N}{2})^2-\frac{N}{2}}{2}+4\right)$} for $N\geq8$, even. The index is therefore:
$\frac{(\frac{N}{2})^2-\frac{N}{2}}{2}+4$\\
\section{Spectra for small $N$}
The following are spectra of extended algebra for several low N.\\
\begin{table}[!htb]
\begin{tabular}{cc}
N=3\\
\begin{tabular}{|c|c|c|}\hline
$\chi^{0,0}_{0}*\theta^{(6)}_0~~0$&$\chi^{0,0}_2*\theta^{(6)}_0~~\frac{4}{5}$&$\chi^{0,0}_{4}*\theta^{(6)}_0~~\frac{2}{5}$\\\hline
$\chi^{1,0}_{0}*\theta^{(6)}_0~~\frac{4}{3}$&$\chi^{1,0}_2*\theta^{(6)}_0~~\frac{2}{15}$&$\chi^{1,0}_{4}*\theta^{(6)}_0~~\frac{11}{15}$\\\hline
$\chi^{2,0}_{0}*\theta^{(6)}_0~~\frac{4}{3}$&$\chi^{2,0}_2*\theta^{(6)}_0~~\frac{2}{15}$&$\chi^{2,0}_{4}*\theta^{(6)}_0~~\frac{11}{15}$\\\hline
$\chi^{0,0}_{0}*\theta^{(6)}_{\pm3}~~\frac{3}{8}$&$\chi^{0,0}_2*\theta^{(6)}_{\pm3}~~\frac{47}{40}$&$\chi^{0,0}_{4}*\theta^{(6)}_{\pm3}~~\frac{31}{40}$\\\hline
$\chi^{1,0}_{0}*\theta^{(6)}_{\pm3}~~\frac{41}{24}$&$\chi^{1,0}_2*\theta^{(6)}_{\pm3}~~\frac{61}{120}$&$\chi^{1,0}_{4}*\theta^{(6)}_{\pm3}~~\frac{133}{120}$\\\hline
$\chi^{2,0}_{0}*\theta^{(6)}_{\pm3}~~\frac{41}{24}$&$\chi^{2,0}_2*\theta^{(6)}_{\pm3}~~\frac{61}{120}$&$\chi^{2,0}_{4}*\theta^{(6)}_{\pm3}~~\frac{133}{120}$\\\hline
$\chi^{0,0}_{0}*\theta^{(6)}_6~~\frac{3}{2}$&$\chi^{0,0}_2*\theta^{(6)}_6~~\frac{23}{10}$&$\chi^{0,0}_{4}*\theta^{(6)}_6~~\frac{19}{10}$\\\hline
$\chi^{1,0}_{0}*\theta^{(6)}_6~~\frac{17}{6}$&$\chi^{1,0}_2*\theta^{(6)}_6~~\frac{49}{30}$&$\chi^{1,0}_{4}*\theta^{(6)}_6~~\frac{67}{30}$\\\hline
$\chi^{2,0}_{0}*\theta^{(6)}_6~~\frac{17}{6}$&$\chi^{2,0}_2*\theta^{(6)}_6~~\frac{49}{30}$&$\chi^{2,0}_{4}*\theta^{(6)}_6~~\frac{67}{30}$\\\hline
\end{tabular}\\
\end{tabular}
\end{table}
\begin{table}[!htb]
\begin{tabular}{cc}
N=4\\
\begin{tabular}{|c|c|c|c|}\hline
$\chi^{0,0}_{0}*\theta^{(2)}_0~~0$&$\chi^{0,0}_{0,2}*\theta^{(2)}_0~~\frac{2}{3}$&$\chi^{0,0}_{2,0}*\theta^{(2)}_0~~\frac{2}{3}$&$\chi^{0,0}_{2,2}*\theta^{(2)}_0~~\frac{1}{3}$\\\hline
$\chi^{1,0}_{1,1}*\theta^{(2)}_0~~\frac{1}{8}$&$\chi^{0,1}_{1,1}*\theta^{(2)}_1~~\frac{1}{4}$&$\chi^{2,1}_{1,1}*\theta^{(2)}_1~~\frac{3}{4}$&\\\hline
$\chi^{1,1}_{0}*\theta^{(2)}_0~~\frac{7}{8}$&$\chi^{1,1}_{0,2}*\theta^{(2)}_0~~\frac{13}{24}$&$\chi^{1,1}_{2,0}*\theta^{(2)}_1~~\frac{13}{24}$&$\chi^{1,1}_{2,2}*\theta^{(2)}_1~~\frac{5}{24}$\\\hline
$\chi^{2,0}_{0}*\theta^{(2)}_0~~\frac{1}{2}$&$\chi^{2,0}_{0,2}*\theta^{(2)}_0~~\frac{1}{6}$&$\chi^{2,0}_{2,0}*\theta^{(2)}_0~~\frac{1}{6}$&$\chi^{2,0}_{2,2}*\theta^{(2)}_0~~\frac{5}{6}$\\\hline
$\chi^{3,0}_{1,1}*\theta^{(2)}_0~~\frac{1}{8}$&&&\\\hline
$\chi^{3,1}_{0}*\theta^{(2)}_0~~\frac{7}{8}$&$\chi^{3,1}_{0,2}*\theta^{(2)}_0~~\frac{13}{24}$&$\chi^{3,1}_{2,0}*\theta^{(2)}_1~~\frac{13}{24}$&$\chi^{3,1}_{2,2}*\theta^{(2)}_1~~\frac{5}{24}$\\\hline
\end{tabular}
\end{tabular}
\end{table}\\
To save the space, for $N=5$ part of the spectrum is omitted, e.g. the fields $\chi^{l,0}_\nu$ and $\chi^{(N-l),0}_\nu$ have the same
dimensions and similar fusions.

\footnotesize
\begin{center}N=5\end{center}
\begin{tabular}[!htb]{|@{}c@{}|@{}c@{}|@{}c@{}|@{}c@{}|@{}c@{}|@{}c@{}|}\hline
$\chi^{0,0}_{(0,0)}*\theta^{(10)}_0~~0$&$\chi^{0,0}_{(1,0)}*\theta^{(10)}_0~~\frac{5}{7}$&$\chi^{0,0}_{(0,2)}*\theta^{(10)}_0~~\frac{4}{7}$
&$\chi^{0,0}_{(2,0)}*\theta^{(10)}_0~~\frac{2}{7}$&$\chi^{0,0}_{(1,2)}*\theta^{(10)}_0~~\frac{1}{7}$&$\chi^{0,0}_{(0,4)}*\theta^{(10)}_0~~\frac{6}{7}$\\\hline
$\chi^{1,0}_{(0,0)}*\theta^{(10)}_0~~\frac{2}{5}$&$\chi^{1,0}_{(1,0)}*\theta^{(10)}_0~~\frac{4}{35}$&$\chi^{1,0}_{(0,2)}*\theta^{(10)}_0~~\frac{34}{35}$
&$\chi^{1,0}_{(2,0)}*\theta^{(10)}_0~~\frac{24}{35}$&$\chi^{1,0}_{(1,2)}*\theta^{(10)}_0~~\frac{19}{35}$&$\chi^{1,0}_{(0,4)}*\theta^{(10)}_0~~\frac{9}{35}$\\\hline
$\chi^{2,0}_{(0,0)}*\theta^{(10)}_0~~\frac{3}{5}$&$\chi^{2,0}_{(1,0)}*\theta^{(10)}_0~~\frac{11}{35}$&$\chi^{2,0}_{(0,2)}*\theta^{(10)}_0~~\frac{6}{35}$
&$\chi^{2,0}_{(2,0)}*\theta^{(10)}_0~~\frac{31}{35}$&$\chi^{2,0}_{(1,2)}*\theta^{(10)}_0~~\frac{26}{35}$&$\chi^{2,0}_{(0,4)}*\theta^{(10)}_0~~\frac{16}{35}$\\\hline
$\chi^{0,0}_{(0,0)}*\theta^{(10)}_{\pm5}~~\frac{5}{8}$&$\chi^{0,0}_{(1,0)}*\theta^{(10)}_{\pm5}~~\frac{75}{56}$&$\chi^{0,0}_{(0,2)}*\theta^{(10)}_{\pm5}~~\frac{67}{56}$
&$\chi^{0,0}_{(2,0)}*\theta^{(10)}_{\pm5}~~\frac{51}{56}$&$\chi^{0,0}_{(1,2)}*\theta^{(10)}_{\pm5}~~\frac{1}{7}$&$\chi^{0,0}_{(0,4)}*\theta^{(10)}_{\pm5}~~\frac{83}{56}$\\\hline
$\chi^{1,0}_{(0,0)}*\theta^{(10)}_{\pm5}~~\frac{41}{40}$&$\chi^{1,0}_{(1,0)}*\theta^{(10)}_{\pm5}~~\frac{207}{280}$&$\chi^{1,0}_{(0,2)}*\theta^{(10)}_{\pm5}~~\frac{167}{280}$
&$\chi^{1,0}_{(2,0)}*\theta^{(10)}_{\pm5}~~\frac{367}{280}$&$\chi^{1,0}_{(1,2)}*\theta^{(10)}_{\pm5}~~\frac{327}{280}$&$\chi^{1,0}_{(0,4)}*\theta^{(10)}_{\pm5}~~\frac{249}{280}$\\\hline
$\chi^{2,0}_{(0,0)}*\theta^{(10)}_{\pm5}~~\frac{49}{40}$&$\chi^{2,0}_{(1,0)}*\theta^{(10)}_{\pm5}~~\frac{263}{280}$&$\chi^{2,0}_{(0,2)}*\theta^{(10)}_{\pm5}~~\frac{223}{280}$
&$\chi^{2,0}_{(2,0)}*\theta^{(10)}_{\pm5}~~\frac{143}{280}$&$\chi^{2,0}_{(1,2)}*\theta^{(10)}_{\pm5}~~\frac{383}{280}$&$\chi^{2,0}_{(0,4)}*\theta^{(10)}_{\pm5}~~\frac{303}{280}$\\\hline
$\chi^{0,0}_{(0,0)}*\theta^{(10)}_{10}~~\frac{5}{2}$&$\chi^{0,0}_{(1,0)}*\theta^{(10)}_{10}~~\frac{45}{14}$&$\chi^{0,0}_{(0,2)}*\theta^{(10)}_{10}~~\frac{43}{14}$
&$\chi^{0,0}_{(2,0)}*\theta^{(10)}_{10}~~\frac{39}{14}$&$\chi^{0,0}_{(1,2)}*\theta^{(10)}_{10}~~\frac{37}{14}$&$\chi^{0,0}_{(0,4)}*\theta^{(10)}_{10}~~\frac{47}{14}$\\\hline
$\chi^{1,0}_{(0,0)}*\theta^{(10)}_{10}~~\frac{29}{10}$&$\chi^{1,0}_{(1,0)}*\theta^{(10)}_{10}~~\frac{183}{70}$&$\chi^{1,0}_{(0,2)}*\theta^{(10)}_{10}~~\frac{173}{70}$
&$\chi^{1,0}_{(2,0)}*\theta^{(10)}_{10}~~\frac{223}{70}$&$\chi^{1,0}_{(1,2)}*\theta^{(10)}_{10}~~\frac{213}{70}$&$\chi^{1,0}_{(0,4)}*\theta^{(10)}_{10}~~\frac{193}{70}$\\\hline
$\chi^{2,0}_{(0,0)}*\theta^{(10)}_{10}~~\frac{31}{10}$&$\chi^{2,0}_{(1,0)}*\theta^{(10)}_{10}~~\frac{197}{70}$&$\chi^{2,0}_{(0,2)}*\theta^{(10)}_{10}~~\frac{187}{70}$
&$\chi^{2,0}_{(2,0)}*\theta^{(10)}_{10}~~\frac{237}{70}$&$\chi^{2,0}_{(1,2)}*\theta^{(10)}_{10}~~\frac{227}{70}$&$\chi^{2,0}_{(0,4)}*\theta^{(10)}_{10}~~\frac{207}{70}$\\\hline
\end{tabular}
\normalsize
\section{Conclusions}
In this paper generalized pseudoaffine theories are defined. They come instead of pseudoaffine theories when
forbidden $q$ is considered. These theories are obtained as a product of generalized parafermions with the
generalized bosons. A specific example of $q=2~Z_N$ parafermions is considered. The generalized pseudoaffine
theory that is obtained has several remarkable properties: number of fields in the spectrum is an integer multiple of that for $SU(2)_N$, for
every field $\phi_i$ of $SU(2)_N$ there is a field $\hat{\phi}_i$ with dimension $q\Delta(\phi_i)$ and
fusion coefficient: $N^k_{ij}=\hat{N}^k_{ij}$. The central charge of the generalized pseudoaffine theory
is $qc_{SU(2)_N}\rm{mod}~4$. In fact by multiplying the parafermions with different generalized bosons (as
explained in the introduction) an infinite hierarchy of such theories is obtained.
\section*{Acknowledgements}
We wish to thank the referee for valueble and important comments.
 

\begin{thebibliography} {33}
\bibitem{gepner92}{D.~Gepner, ``Foundations of rational quantum field theory. 1,'' hep-th/9211100.}
\bibitem{gepner98}{E.~Baver, D.~Gepner and U.~G$\ddot{\rm{u}}$rsoy, ``On conformal field theories at fractional 
levels,'' Nucl.\ Phys.\ {\bf B557}, 505 (1999) [hep-th/9811100].}
\bibitem{gepner99a}{D.~Gepner, ``On new conformal field theories with affine fusion rules,'' Nucl.\ Phys.\ 
{\bf B561}, 467 (1999) [hep-th/9901022].}
\bibitem{gepner99}{E.~Baver, D.~Gepner and U.~G$\ddot{\rm{u}}$rsoy, ``Realizations of pseudobosonic theories with non-diagonal 
automorphisms,'' Nucl.\ Phys.\ {\bf B561}, 473 (1999) [hep-th/9905164].}
\bibitem{ZF}{V.~A.~Fateev and A.~B.~Zamolodchikov, ``Parafermionic Currents In The Two-Dimensional Conformal 
Quantum Field Theory And Selfdual Critical Points In Z(N) Invariant Statistical Systems,'' Zh. Eksp. Theor. Fiz
89(1985)215 translated in Sov.\ Phys.\ JETP{\bf 62}, 215 (1985).}
\bibitem{ZF2}{V.~A.~Fateev and A.~B.~Zamolodchikov, ``Representations Of The Algebra Of 'Parafermion Currents' 
Of Spin 4/3 In Two-Dimensional Conformal Field Theory. Minimal Models And The Tricritical Potts Z(3) Model,''
Theor.\ Math.\ Phys.\ {\bf 71}, 451 (1987).}
\bibitem{GQ}{D.~Gepner and Z.~Qiu, ``Modular Invariant Partition Functions For Parafermionic Field Theories,''
Nucl.\ Phys.\ {\bf B285}, 423 (1987).}
\bibitem{G87}{D.~Gepner, ``New Conformal Field Theories Associated With Lie Algebras And Their Partition 
Functions,'' Nucl.\ Phys.\ {\bf B290}, 10 (1987).}
\bibitem{schwimmer}{P.~Goddard and A.~Schwimmer, ``Unitary Construction Of Extended Conformal Algebras,''
Phys.\ Lett.\ {\bf B206}, 62 (1988).}
\bibitem{CFT}{Philippe Di Francesco, Pierre Mathieu, David S$\acute{e}$n$\acute{e}$chal, "Conformal Field Theory",
Springer Verlag 1997}
\bibitem{Rav}{F.~Ravanini,``On the possibility of Z(N) exotic supersymmetry in two-dimensional conformal field 
theory,''Int.\ J.\ Mod.\ Phys.\ {\bf A7}, 4949 (1992)[hep-th/9109057].}
\end{thebibliography}
\end{document}